\documentclass[%
 reprint,
nofootinbib,
 amsmath,amssymb,
 aps,
pre,
showpacs
]{revtex4-1}

\usepackage{graphicx}
\usepackage{dcolumn}
\usepackage{bm}
\usepackage{natbib}

\def\longrightharpoonup{\relbar\joinrel\rightharpoonup}
\def\longleftharpoondown{\leftharpoondown\joinrel\relbar}

\def\longrightleftharpoons{
  \mathop{
    \vcenter{
      \hbox{
      \ooalign{
        \raise1pt\hbox{$\longrightharpoonup\joinrel$}\crcr
	  \lower1pt\hbox{$\longleftharpoondown\joinrel$}
	  }
      }
    }
  }
}

\begin{document}


\title{Thermal Fluctuation Statistics in a Molecular Motor \\ Described by a Multidimensional Master Equation}

\author{K. J.~Challis}
\author{M. W.~Jack}%

\affiliation{%
 Scion, 49 Sala Street, Private Bag 3020, Rotorua 3046, New Zealand
}%




\date{\today}

\begin{abstract}
We present a theoretical investigation of thermal fluctuation statistics in a molecular motor.  Energy transfer in the motor is described using a multidimensional discrete master equation with nearest-neighbor hopping.  In this theory, energy transfer leads to statistical correlations between thermal fluctuations in different degrees of freedom.  For long times, the energy transfer is a multivariate diffusion process with a constant drift and diffusion.  The fluctuations and drift align in the strong-coupling limit enabling a one-dimensional description along the coupled coordinate.  We derive formal expressions for the probability distribution and simulate single trajectories of the system in the near and far from equilibrium limits both for strong and weak coupling.  Our results show that the hopping statistics provide an opportunity to distinguish different operating regimes.

\end{abstract}


\maketitle

\section{Introduction}

Biological systems use specialized motor proteins to convert energy from one form to another \cite{Howard2001, Chowdhury2013}.  These nanoscale devices operate in an environment dominated by thermal fluctuations and it has been suggested that they harness thermal fluctuations to perform their tasks effectively \cite{Astumian2007, Ishii2008, Beausang2013}.  Single-molecule experiments can probe molecular motor fluctuation statistics \cite{Nishiyama2002, Yildiz2003, Shimabukuro2003, Toba2006} and this provides new opportunities to elucidate molecular motor mechanisms \cite{Svoboda1994, Shaevitz2005}. 

The intrinsic stochasticity of molecular motor operation has been formalized in a range of Brownian motion-based theories \cite{Magnasco1994, Fisher1999, Wang2008, Lipowsky2009, Toyabe2010, Chowdhury2013}. One approach has been to describe molecular motors in terms of Brownian motion on a tilted periodic potential, either via continuous Fokker-Planck equations \cite{Magnasco1994, Keller2000, Golubeva2012, VandenBroeck2012, ChallisarXiv} or discrete hopping models \cite{Fisher2005, Lau2007, Schmiedl2008, Zhang2011, Bameta2013}.  In the long-time limit, the system can be described as a Gaussian process with an effective drift and diffusion \cite{Lindner2001, Reimann2002, Pavliotis2005}.  The drift quantifies the average rate of transport in the system and the diffusion quantifies the thermal fluctuations.  

In this paper, we present a systematic theoretical treatment of the thermal fluctuation statistics of a molecular motor.  Our approach is based on a multidimensional discrete master equation with nearest-neighbor hopping.  This enables us to clarify fundamental aspects of the thermal fluctuation statistics for coupled processes.  The particular advantages of our approach are (i) it describes more than one degree of freedom making coupling between degrees of freedom explicit, (ii) it includes loss processes providing access to both the strong and weak coupling regimes, and (iii) it is tractable analytically yielding formal results that build our intuition for these systems.  The key result is that energy transfer leads to statistical correlations between thermal fluctuations in different degrees of freedom.  These correlations depend on the operating regime of the system and are accessible via the diffusion matrix, discrete hopping statistics, and single trajectories.

This paper is organized as follows. In Sec.\ \ref{sec:background} we present our theoretical description of energy transfer in a molecular motor.  In Sec.\ \ref{sec:long} we consider the long-time Gaussian limit of the system and describe signatures of energy transfer in the drift vector and diffusion matrix.  In Sec.\ \ref{sec:hop} we derive the waiting time and formal expressions for the probability distribution in the strong and weak coupling regimes both near and far from equilibrium.  In Sec.\ \ref{sec:single} we simulate single trajectories of the system.  We conclude in Sec.\ \ref{sec:conc}.

\section{Multidimensional Master Equation \label{sec:background}}

Energy transfer in a molecular motor has been described in terms of Brownian motion on a multidimensional free-energy landscape \cite{Magnasco1994, Keller2000, Fisher2005}. In this theory, each degree of freedom is a generalized coordinate capturing the main conformal motions of the motor and representing displacements either in real space or along reaction coordinates \cite{Kramers1940}.  In the limit of deep potential wells, the system is confined to (meta-)stable states of the potential and it is intuitive that the continuous theory can be approximated by a discrete equation describing thermally activated hopping between potential wells.  We consider the model potential \cite{Magnasco1994, Keller2000}
\begin{equation}
V(\bm r) = V_{\bm 0}(\bm r) - \bm f \cdot \bm r,
\end{equation}
where $\bm r$ is the position, the potential $V_{\bm 0}(\bm r)= V_{\bm 0}(\bm r +a_j \hat{\bm r}_j)= V_{\bm 0}(\bm r +\bm a)$ is periodic with period $\bm a$, and $\bm f$ represents macroscopic thermodynamic forces that drive the system out of thermal equilibrium.  For long times, we write the master equation \cite{Challis2013}
\begin{equation}
\frac{d p_{\bm n}(t)}{dt}=\sum_{{\bm n}'} \left[\kappa^{\bm f}_{{\bm n}-{\bm n}'}p_{\bm n'}(t)-\kappa^{\bm f}_{{\bm n}'-{\bm n}} p_{\bm n}(t)\right] \label{master_eqn},
\end{equation}
where $p_{\bm n}(t)$ is the probability of state $\bm n$ being occupied, and $\bm n$ and $\bm n'$ are vectors of integers.  The transition rates $\kappa_{\bm n}^{\bm f}$ satisfy $\sum_{{\bm n}}\kappa^{\bm f}_{{\bm n}}=0$, and we consider the summation to be over nearest neighbor states only.  To lowest order, the dependence on the thermodynamic force $\bm f$ is assumed to take the form \cite{Fisher2005, VandenBroeck2012, Challis2013}
\begin{eqnarray}
\kappa^{\bm f}_{{\bm n}}& = &e^{\alpha_{\bm n}{\bm f}\cdot {\bm A}{\bm n}/k_{B}T}\kappa^{\bm 0}_{{\bm n}},\label{rate}
\end{eqnarray}
where $\kappa_{\bm n}^{\bm 0}=\kappa_{-\bm n}^{\bm 0}$ are the transition rates at equilibrium ($\bm f = \bm 0$), $k_B$ is the Boltzmann constant, and $T$ is the temperature.  The matrix $\bm A$ is diagonal with $A_{jj}=a_j$.  The loading coefficients $\alpha_{\bm n}$ satisfy generalized detailed balance, i.e., $\alpha_{-\bm n}=1-\alpha_{\bm n}$ and $0 \leq \alpha_{\bm n}\leq 1$ \cite{Fisher1999, Seifert2010}.  Energy transfer between degrees of freedom is possible when the transition rates $\kappa_{\bm n}^{\bm 0}$ are non-vanishing for $n_j\neq0$ in more than one coordinate and transport occurs along coupled coordinates \cite{ChallisarXiv}.
 
The master equation (\ref{master_eqn}) can be solved analytically by transforming to the diagonal form
\begin{equation}
\frac{dc_{\bm k}(t)}{dt}=-\lambda_{\bm k} c_{\bm k}(t) \label{diagonal},
\end{equation}
where the eigenstates are
\begin{equation}
c_{\bm k}(t)=\sum_{\bm n} p_{\bm n}(t) e^{-i{\bm k}\cdot \bm A \bm n},
\label{characteristic_function_discrete}
\end{equation}
 and the eigenvalues are
\begin{eqnarray}
\lambda_{\bm k} & = &   \sum_{{\bm n} \in {\rm for}}4\kappa^{\bm 0}_{\bm n}  G_{\bm n}\left(\frac{\bm X\cdot \bm n}{2}\right) \sin\left(\frac{{\bm k}\cdot \bm A\bm n}{2}\right) \nonumber \\
& & \times \sin \left( \frac{\bm k \cdot \bm A \bm n}{2} +\frac{i \bm X  \cdot \bm n}{2}\right).
\label{eigenvalues_general}
\end{eqnarray}
In Eq.\ (\ref{eigenvalues_general}), we have identified the unitless generalized thermodynamic force $\bm X = \bm A\bm f /k_BT$ \cite{ChallisarXiv} and defined the loading function
\begin{equation}
G_{\bm n}(x) = e^{(2\alpha_{\bm n}-1)x }.
\end{equation}
The summation over $\bm n$ in Eq.\ (\ref{eigenvalues_general}) is taken over forward rates only, i.e., over one of $\bm n$ or $-\bm n$, not both.  The analytic solution to Eq.\ (\ref{diagonal}) is
\begin{equation}
c_{\bm k}(t)=c_{\bm k}(0)e^{-\lambda_{\bm k}t}. 
\label{characteristic_function_solution}
\end{equation}

\section{Long-time Limit \label{sec:long}}

In the long-time limit, Brownian motion on a tilted periodic potential can be described by a Guassian process with an effective drift and diffusion \cite{Lindner2001, Reimann2002, Pavliotis2005}.  This makes drift and diffusion key predictors of Brownian motion-based theories of molecular motors.  In the multidimensional case, the drift and diffusion contain signatures of the energy transfer.

\subsection{Drift and Diffusion}

The effective drift vector $\bm v$ in the long-time limit is given by
\begin{equation}
v_j= \lim_{t\rightarrow\infty} \frac{\langle n_j\rangle}{t},
\end{equation}
and the effective diffusion matrix $D$ is
\begin{equation}
D_{jj'} =  \lim_{t\rightarrow \infty} \frac{\langle n_j n_{j'}\rangle -\langle n_j\rangle \langle n_{j'}\rangle}{2t}.
\label{diff_matrix}
\end{equation} 
For the master equation (\ref{master_eqn}), the drift is equal to the average rate of transport and is given by
\begin{eqnarray}
v_j &  = & \frac{d\langle n_j \rangle}{dt}\\
& = & \sum_{\bm n \in for}   2\kappa_{\bm n}^{\bm 0} n_j G_{\bm n}\left(\frac{\bm X\cdot \bm n}{2}\right)\sinh \left( \frac{\bm X \cdot \bm n}{2}\right). \label{velocity_multi_dim}
\end{eqnarray}
Equation (\ref{velocity_multi_dim}) shows that when transitions occur simultaneously in more than one degree of freedom (i.e., when $\kappa_{\bm n}^{\bm 0}$ is non-vanishing for $n_j\neq0$ in more than one coordinate), the force in one degree of freedom can drive drift in another.  This drift can occur even against an opposing force representing energy transfer between degrees of freedom.  Near equilibrium, where $|X_{j}|\ll 1$, the drift becomes 
\begin{eqnarray}
v_j \approx \sum_{\bm n\in for} \sum_{j'} \kappa_{\bm n}^{\bm 0} n_j n_{j'} X_{j'}.
\label{drift_eq}
\end{eqnarray}
Equation (\ref{drift_eq}) is a linear force-flux relation satisfying the Onsager relations \cite{ChallisarXiv}.

The diffusion matrix for the master equation (\ref{master_eqn}) is related to the covariance matrix $\sigma$ \cite{Gardiner2009} and is given by
\begin{eqnarray}
D_{jj'} &  = & \frac{d(\langle n_j n_{j'}\rangle - \langle n_j\rangle \langle n_{j'}\rangle)}{2dt} = \frac{1}{2}\frac{d\sigma_{jj'}}{dt}  \\
& = & \sum_{\bm n \in for} 2\kappa_{\bm n}^{\bm 0} n_j n_{j'} G_{\bm n}\left(\frac{\bm X\cdot \bm n}{2}\right) \cosh \left( \frac{\bm X \cdot \bm n}{2}\right).
 \label{gamma_multi_dim}
\end{eqnarray}
The diffusion matrix provides insights into the thermal fluctuations of the system.  For coupled transitions occuring simultaneously in more than one coordinate, $D_{jj'}$ can be non-zero for $j\neq j'$.  This describes statistical correlations between fluctuations in different degrees of freedom.  Near equilibrium, thermal diffusion still occurs and the diffusion matrix becomes
\begin{eqnarray}
D&\approx&\sum_{{\bm n} \in {\rm for}} \kappa^{\bm 0}_{\bm n}n_{j}n_{j'}.
\label{sigma_eq}
\end{eqnarray} 

Using the drift (\ref{velocity_multi_dim}) and diffusion (\ref{gamma_multi_dim}), we find that
\begin{eqnarray}
\frac{\partial v_j}{\partial X_{j'}} & = &  D_{jj'}  \label{Einstein_rel} \\
& & +\sum_{\bm n \in {\rm for}} 2\kappa_{\bm n}^{\bm 0} n_j n_{j'} (\alpha_{\bm n}-1/2) G_{\bm n} \left(\frac{\bm X \cdot \bm n}{2}\right) \sinh \left(\frac{\bm X \cdot \bm n}{2}\right). \nonumber
\end{eqnarray} 
Equation (\ref{Einstein_rel}) takes the form of a generalized Einstein relation and, as expected, the second term on the right-hand side vanishes when $\alpha_{\bm n}=1/2$, recovering the equilibrium result \cite{Seifert2010, Speck2006}.

\subsection{Gaussian Approximation \label{sec:diffusion}}

The master equation (\ref{master_eqn}) can be approximated for long times by a continuous diffusion equation with a constant effective drift vector and constant effective diffusion matrix, as follows. The eigensates $c_{\bm k}(t)$ decay according to the evolution equation (\ref{diagonal}) and, for long times, only the lowest eigensates remain populated and contribute to the system dynamics.  As described in Appendix  \ref{sec:cumulant}, we approximate the evolution equation by replacing $\lambda_{\bm k}$ by its second-order Taylor series around the origin, i.e.,
\begin{equation}
\lambda_{\bm k}\approx-i\bm A{\bm k}\cdot{\bm v} - 2(\bm A{\bm k})^{T}D {\bm A\bm k},
\label{lambda_approx}
\end{equation}
where $\bm k$ is taken as a column vector.  The probability coefficients $p_{\bm n}(t)$ can be determined from the eigenstates $c_{\bm k}(t)$ by inverting Eq.\ (\ref{characteristic_function_discrete}), i.e., we define the continuous probability density
\begin{equation}
{\mathcal P}({\bm s},t) = \frac{1}{(2\pi)^d}\int d{\bm k}\  c_{\bm k}(t) e^{i{\bm k}\cdot \bm A{\bm s}},
\label{characteristic_function_continuous}
\end{equation} 
where $s$ is a unitless continuous variable.  The equation of motion for ${\mathcal P}({\bm s},t)$ is 
\begin{equation}
\frac{\partial {\mathcal P}({\bm s},t)}{\partial t}=\left(-{\bm v}\cdot\nabla_{\bm s} +\sum_{jj'} 2D_{jj'} \frac{\partial^{2}}{\partial s_{j}\partial s_{j'}}\right) {\mathcal P}({\bm s},t),
\label{diffusion}
\end{equation}
which describes a multivariate diffusion process where the drift vector and diffusion matrix depend on the periodic potential, via $\kappa_{\bm n}^{\bm 0}$ and $\alpha_{\bm n}$, and the force $\bm f$ according to Eqs.\ (\ref{velocity_multi_dim}) and (\ref{gamma_multi_dim}).  

If the system is initially described by a Dirac delta function at the origin, the diffusion equation (\ref{diffusion}) can be solved analytically to yield
\begin{equation}
{\cal P}_{ss} (\bm s,t) =  \sqrt{\frac{8|D|}{(2\pi)^d t}}  \exp \left( -\frac{1}{8t} (\bm s-\bm v t)^{T}  D^{-1} (\bm s-\bm v t)\right). \label{Gaussian}
\end{equation}
In the long-time limit, the steady state is independent of the initial condition so the Gaussian (\ref{Gaussian}) provides a good approximate description of the system. This means that energy transfer in the steady state can be interpreted as a Gaussian process where, in general, the principal axis of the diffusion matrix is not aligned with the drift vector. 

It is straightforward to write down the Ito stochastic differential equation for Eq.\ (\ref{diffusion}) and derive the two-time correlation function in the long-time limit \cite{Gardiner2009}:
\begin{eqnarray}
\langle s_j(t) , s_{j'} (t')\rangle &  = & \langle s_j(t)  s_{j'} (t')\rangle-\langle s_j(t)\rangle \langle   s_{j'} (t')\rangle \\
& = & 4D_{jj'} \min (t,t').
\end{eqnarray}

\subsection{Two Dimensions \label{sec:2D}}

To explicitly interprete the transfer of energy between degrees of freedom, we consider the case with just two dimensions.  Labeling the two coordinates $x$ and $y$, the eigenvalues (\ref{eigenvalues_general}) are 
\begin{eqnarray}
\lambda_{(k_{x},k_{y})}&=&4\kappa^{\bm 0}_{(1,0)}G_{(1,0)}(X_x/2)\sin(k_x a_x/2 )\sin(k_x a_x/2+i X_x/2) \nonumber \\
& & +4\kappa^{\bm 0}_{(0,1)}G_{(0,1)}(X_y/2)\sin(k_y a_y/2)\sin ( k_y a_y/2+iX_y/2) \nonumber \\
&&+4\kappa^{\bm 0}_{(1,1)}G_{(1,1)}(X_z/2)\sin(k_{x}a_{x}/2+k_{y}a_{y}/2) \nonumber \\
& & \times \sin( k_{x}a_{x}/2+k_{y}a_{y}/2+iX_z/2),
\label{eigenvalues_2D}
\end{eqnarray}
where we have identified $X_z = X_x+X_y$ as the thermodynamic force along the coupled coordinate, and assumed that coupling between degrees of freedom is preferential in the $(1,1)$ direction so that $\kappa^{\bm 0}_{(1,-1)}$ is negligible.\footnote{The effect of the competing orthogonal coupling transition, described in our formalism by $\kappa_{(1,-1)}^{\bm 0}$, has been considered by other authors \cite{Golubeva2012}.}  The uncoupled transition rates $\kappa^{\bm 0}_{(0,1)}$ and $\kappa^{\bm 0}_{(1,0)}$ represent {\em leak processes} that bypass the coupling mechanism and weaken the coupling between the $x$ and $y$ degrees of freedom \cite{Lems2003}.  The drift (\ref{velocity_multi_dim}) becomes
\begin{eqnarray}
v_{x}
&=&2\kappa^{\bm 0}_{(1,0)} G_{(1,0)}(X_x/2)\sinh(X_{x}/2) \nonumber \\
& & +2\kappa^{\bm 0}_{(1,1)} G_{(1,1)}(X_z/2)\sinh(X_z/2)\label{drift1}\\
v_{y}
&=&2\kappa^{\bm 0}_{(0,1)}G_{(0,1)}(X_y/2)\sinh(X_y/2) \nonumber \\
& & +2\kappa^{\bm 0}_{(1,1)}G_{(1,1)}(X_z/2)\sinh(X_z/2)\label{drift2}
\end{eqnarray}
and the diffusion (\ref{diff_matrix}) becomes
\begin{eqnarray}
D_{xx}&=&\kappa^{\bm 0}_{(1,0)}G_{(1,0)}(X_x/2)\cosh(X_x/2) \nonumber \\
&&+\kappa^{\bm 0}_{(1,1)}G_{(1,1)}(X_z/2)\cosh(X_z/2) \\
D_{yy}&=&\kappa^{\bm 0}_{(0,1)}G_{(0,1)}(X_y/2)\cosh(X_y/2) \nonumber \\
& &+\kappa^{\bm 0}_{(1,1)}G_{(1,1)}(X_z/2)\cosh(X_z/2) \\
D_{xy}&=&D_{yx}=\kappa^{\bm 0}_{(1,1)}G_{(1,1)}(X_z/2)\cosh(X_z/2). \label{diff_offdiag}
\end{eqnarray}

For $\kappa_{(1,1)}^{\bm 0}= 0$, coupling between degrees of freedom can not occur.  In this case, the coordinates decouple so that $v_x$ and $D_{xx}$ depend only on $X_x$,  $v_y$ and $D_{yy}$ depend only on $X_y$, and the correlation term $D_{xy}$ vanishes.  In contrast, for $\kappa_{(1,1)}^{\bm 0}\neq 0$, the thermodynamic force in one degree of freedom can drive drift in the other.  For example, $v_x$ depends on $X_y$ via the coupled force $X_z=X_x+X_y$.  In addition,  coupling between the degrees of freedom induces correlations between the thermal fluctuations.  This is illustrated in two ways: (i) the force in one degree of freedom can drive fluctuations in the other (e.g., $D_{xx}$ depends on $X_y$), and (ii) the  correlation term $D_{xy}$ can be non-vanishing.  The key result is that energy transfer between degrees of freedom leads to statistical correlations between thermal fluctuations in those degrees of freedom.

\begin{figure}
	\centering
		\includegraphics{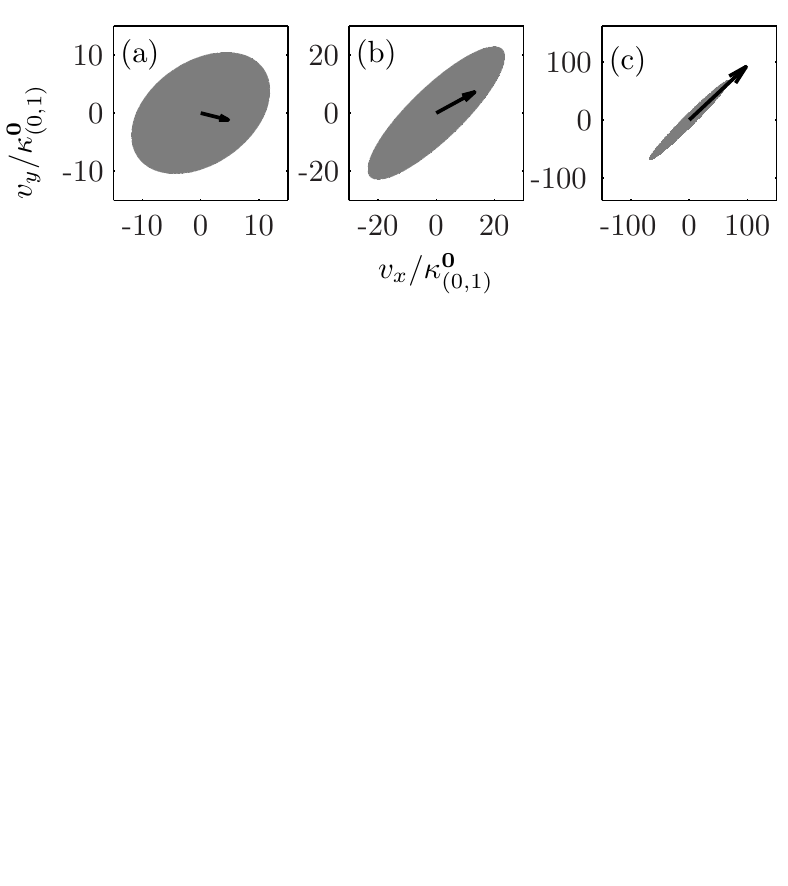}
		\vspace{-6.3cm}
		\caption{Drift vector for (a) $\kappa^{\bm 0}_{(1,1)}= \kappa^{\bm 0}_{(0,1)}$, (b) $\kappa^{\bm 0}_{(1,1)}=10\kappa^{\bm 0}_{(0,1)}$, and (c) $\kappa^{\bm 0}_{(1,1)}=100 \kappa^{\bm 0}_{(0,1)}$.  Other parameters are $X_{x}=3 $, $X_{y}=-2$, $\kappa^{\bm 0}_{(1,0)}= \kappa^{\bm 0}_{(0,1)}$, and $\alpha_{(0,1)}=\alpha_{(1,0)}=\alpha_{(1,1)}=1/2$.  The  ellipse is a shaded contour of the Gaussian probability density (\ref{Gaussian}) centered at the origin.}
	\label{fig:velocity_covariance}
\end{figure}

The energy transfer can be visualized in the $x$-$y$ plane as a drift vector and an elliptical contour of the diffusing Gaussian probability density (\ref{Gaussian}) [centered at the origin for simplicity].  This is shown in Fig.\ \ref{fig:velocity_covariance} for three different values of the coupling strength $\kappa^{\bm 0}_{(1,1)}/\kappa^{\bm 0}_{(0,1)}$.  We consider $X_x>0$ and $X_y<0$ so that transitions in the $x$ coordinate are thermodynamically favourable (spontaneous) and transitions in the $y$ coordinate are thermodynamically unfavourable (nonspontaneous).  We also choose $X_z=X_x+X_y>0,$ so that transitions in the coupled coordinate $z=x+y$ are thermodynamically favourable (spontaneous).  Figure \ref{fig:velocity_covariance}(a) shows the weak-coupling limit where the coupling strength $\kappa^{\bm 0}_{(1,1)}/\kappa^{\bm 0}_{(0,1)}\lesssim 1$.  In this case, the drift vector in the $y$ coordinate is negative, in the same direction as the force in that coordinate $X_{y}<0$, and the diffusion ellipse is roughly symmetric because the fluctuations are weakly correlated.  Figure \ref{fig:velocity_covariance}(b) shows moderate coupling where $\kappa^{\bm 0}_{(1,1)}/\kappa^{\bm 0}_{(0,1)}>1$. In this case, the drift vector points in the positive $y$ direction, opposing the $X_y<0$ force, and the diffusion ellipse becomes elongated as the coupling leads to fluctuations along the coupled $z$ coordinate.  Figure \ref{fig:velocity_covariance}(c) shows strong coupling where $\kappa^{\bm 0}_{(1,1)}/\kappa^{\bm 0}_{(0,1)}\gg1$.  In this case, the drift amplitude is much larger than for weak or moderate coupling, the diffusion ellipse is strongly elongated along the coupled $z$ coordinate, and the drift vector and the principle axis of the diffusion ellipse line up.  For strong coupling, a one-dimensional description in the coupled coordinate is valid, as considered in Sec.\ \ref{sec:strong}.

\section{Hopping Statistics \label{sec:hop}}

The long-time Gaussian approximation described in the previous section provides a physical interpretation of the signatures of energy transfer. However, the master equation (\ref{master_eqn}) describes more detail about the evolution of the system.  In this section, we consider how the signatures of energy transfer manifest in the discrete hopping statistics of the system.  

\subsection{Waiting Time \label{sec:waiting}}

An important measurable quantity is the time delay between hopping events.  This is usually referred to as the waiting time or the dwell time \cite{Fisher1999}.  Assuming  the system initially occupies state $\bm n_{0}$, the probability $p_{\bm n_0}(t)$ of occupying the state $\bm n_0$ at time $t$ evolves according to the master equation (\ref{master_eqn}).  For an infinitesimal $t$, $p_{\bm n}(t)$ is negligible unless $\bm n = \bm n_0$, so Eq.\ (\ref{master_eqn}) can be approximated by 
\begin{equation}
\frac{dp_{\bm n_0}(t)}{dt} \approx \kappa_{\bm 0}^{\bm f} p_{\bm n_0}(t).
\label{approx_me}
\end{equation} 
Integrating Eq.\ (\ref{approx_me}) gives
\begin{equation}
p_{\bm n_0}(t) = p_{\bm n_0}(0) e^{\kappa_{\bm 0}^{\bm f} t}= e^{- t/\tau}=e^{-\Gamma t}.
\label{pni}
\end{equation}
Equation (\ref{pni}) shows that the jump times are exponentially distributed. This is a general characteristic of master-equaton models \cite{Gardiner2009} and has been observed in single-molecule experiments of molecular motors \cite{Yildiz2003, Shimabukuro2003, Sakamoto2008}.  In Eq.\ (\ref{pni}), the decay rate $\Gamma$ depends on the thermodynamic force $\bm X$ according to
\begin{eqnarray}
\Gamma & = &  -\kappa_{\bm 0}^{\bm f} =\sum_{\bm n \neq\bm 0} \kappa_{\bm n}^{\bm f} \\
&  = &  \sum_{\bm n \neq\bm 0, \in for} 2\kappa_{\bm n}^{\bm 0} G_{\bm n}\left( \frac{\bm X\cdot \bm n}{2} \right)\cosh\left(\frac{\bm X \cdot \bm n}{2}\right).
\end{eqnarray}
For the two-dimensional case considered in Sec.\ \ref{sec:2D}, 
\begin{eqnarray}
\Gamma&=&2\kappa^{\bm 0}_{(1,0)} G_{(1,0)}(X_x/2)\cosh( X_x/2)\nonumber \\
&&+2\kappa^{\bm 0}_{(0,1)}G_{(0,1)}(X_y/2)\cosh(X_y/2)\nonumber \\
& & +2\kappa^{\bm 0}_{(1,1)}G_{(1,1)}(X_z/2)\cosh(X_z/2).
\end{eqnarray}
Near equilibrium, the decay rate $\Gamma$ becomes independent of the force $\bm X$. 

\subsection{Hopping Rates}

The transition rates $\kappa_{\bm n}^{\bm f}$ depend on the thermodynamic force according to Eq.\ (\ref{rate}), and the average ratio of forward to backward hops is given by the generalized detailed balance condition
\begin{equation}
\frac{\kappa_{\bm n}^{\bm f}}{\kappa_{-\bm n}^{\bm f}}=e^{(\alpha_{\bm n}+\alpha_{-\bm n}){\bm f}\cdot{\bm A \bm n}/k_{B}T} = e^{{\bm X} \bm n}.
\label{ratio}
\end{equation}
The exponential form of Eq.\ (\ref{ratio}) has been observed experimentally for kinesin \cite{Nishiyama2002}. Equation (\ref{ratio}) shows that the rates of forward and backward hopping are approximately equal near equilibrium, when $|X_j|\ll 1$. The situation is quite different when the system operates far from equlibrium.  When $X_j  \gg 1$, the forward hopping rates are exponentially larger than the backward rates, and when $X_j \ll -1$, the backward hopping rates are exponentially larger than the forward rates.  This dependence on the thermodynamic force leads to different behavior in the near and far from equilibrium regimes, as demonstrated in the following sections.

\subsection{Strong Coupling \label{sec:strong}}

The strong-coupling regime has been considered theoretically by other authors \cite{Magnasco1994, Golubeva2012, VandenBroeck2012}, and results from single-molecule experiments indicate that certain molecular motors operate in this regime \cite{Rondelez2005, Sakamoto2008}.  We consider strong coupling in detail here for completeness, and because it clearly demonstrates the difference between the near and far from equilibrium operating regimes.

In the strong-coupling regime, the leak transition rates $\kappa_{(1,0)}^{\bm 0}$ and $\kappa_{(0,1)}^{\bm 0}$ are negligible compared to the coupled rate $\kappa_{(1,1)}^{\bm 0}$, and all transitions occur simultaneously in $x$ and $y$.  In this case, the system is well described by the one-dimensional master equation in the coupled coordinate $z=x+y$:
\begin{eqnarray}
\frac{d p_{n_z}(t)}{dt}& = &\kappa_{(1,1)}^{\bm 0}\left[e^{\alpha_{(1,1)}X_z}p_{n_{z}-1}(t)+ e^{(\alpha_{(1,1)}-1)X_z}p_{n_z+1}(t)\right] \nonumber \\
& & -\Gamma_z p_{n_z}(t),
\label{master_eqn_z}
\end{eqnarray}
where $\Gamma_{z}= 2\kappa_{(1,1)}^{\bm 0}G_{(1,1)}(X_z/2) \cosh(X_z/2)$. Assuming an initial state $p_{n_z}(0)=\delta_{n_z 0}$, the general analytic solution to Eq.\ (\ref{master_eqn_z}) is
\begin{equation}
p_{n_z}(t)=e^{X_{z}n_z/2} e^{-\Gamma_zt } I_{n_z}\left(2t\kappa_{(1,1)}^{\bm 0} G_{(1,1)}(X_z/2)\right),
\label{soln}
\end{equation}
where $I_n(x)$ is the modified Bessel function finite at the origin \cite{Abramowitz1972}. 

Near equilibrium, i.e., $X_z \ll 1$, the first exponential term in Eq.\ (\ref{soln}) is of order unity and the probability distribution becomes approximately symmetric in $n_z$.  In the long-time limit, the probability distribution takes the Gaussian form
\begin{equation}
p_{n_z}(t) \sim \frac{e^{-n_z^2/2 t\Gamma_z}}{\sqrt{2\pi t\Gamma_z}} ,
\label{soln_long_time}
\end{equation} 
with $\Gamma_z \approx 2\kappa_{(1,1)}^{\bm 0}$. Equation (\ref{soln_long_time}) shows that the probability $p_{n_z=0}(t)$ of the system initially at $n_z=0$ remaining at $n_z=0$ at time $t$ decays as $1/\sqrt{t}$, characteristic of a diffusion process. 

Far from equilibrium, i.e., $X_z \gg1$, the second term on the right-hand side of the master equation (\ref{master_eqn_z}) is negligible.  In that case, $p_{n_z}(t)$ is well described by the Poisson distribution 
\begin{equation}
p_{n_z}(t) = e^{-\Gamma_z t} \frac{(t\Gamma_z)^{n_z}}{n_z !},
\label{poissonz}
\end{equation} 
with $\Gamma_z=\kappa_{(1,1)}^{\bm 0}\exp(\alpha_{(1,1)}X_{z})$.  Equation (\ref{poissonz}) shows that far from equilibrium $p_{n_z=0}(t)$ decays as $\exp(-t \Gamma_z)$.  For long times,  the Poisson distribution (\ref{poissonz}) is well approximated by a Gaussian, consistent with Sec.\ (\ref{sec:diffusion}).

In general for strong coupling, rotating the drift vector and diffusion matrix to the coupled coordinate $z$, and rescaling to $a_z$, the drift is 
\begin{equation}
v_z = 2 \kappa_{(1,1)}^{\bm 0} G_{(1,1)}(X_z/2)\sinh(X_z/2),
\end{equation}
and the diffusion is
\begin{equation}
D_{zz}= \kappa_{(1,1)}^{\bm 0} G_{(1,1)}(X_z/2)\cosh(X_z/2).
\end{equation}
The relative timescales of the drift and diffusion processes can be compared using either the randomness $r$ or the P\'{e}clet number $Pe$ \cite{Lindner2001, Svoboda1994}.  In the coupled coordinate these quantities satisfy the known result \cite{Lindner2001}
\begin{equation}
r = \frac{2 D_{zz}}{v_z}= \frac{2}{Pe}=\coth(X_z/2).  
\label{peclet}
\end{equation}
Near equilibrium,  $r$ is large and $Pe\ll 1$ indicating that diffusion occurs much faster than drift, characteristic of a diffusion process.  Far from equilibrium,  $r$ tends to 1 and $Pe$ tends to 2 indicating that drift and diffusion occur on similar timescales, characteristic of a Poisson process. 

\subsection{Weak Coupling \label{sec:weak}}

The extent to which molecular motors operate outside the strong-coupling regime remains an open question \cite{Chowdhury2013, Ishijima1998, Oosawa2000, Ikezaki2013}. In this section, we consider the weak-coupling regime where the leak transition rates $\kappa_{(1,0)}^{\bm 0}$ and $\kappa_{(0,1)}^{\bm 0}$ are not negligible compared to the coupled rate $\kappa_{(1,1)}^{\bm 0}$, and a one-dimensional description along the coupled coordinate $z$ is insufficient.  We consider the near and far from equilibrium regimes for the case where each degree of freedom is observed independently. This is relevant because, in practise, it is often not straightforward to access all degrees of freedom simultaneously.

A description of the system along a single degree of freedom can be determined by tracing over the inaccessible or unobserved coordinates.  For example, if the multidimensional system is observed only along the $x$ co-ordinate, the effective probability density is
\begin{equation}
p_{n_x}(t)=\sum_{ n_j\neq n_x}p_{\bm n}(t),
\end{equation} 
which evolves according to the one-dimensional master equation
\begin{equation}
\frac{dp_{n_x}(t)}{dt} =\sum_{\bm n'}\kappa_{\bm n + \bm n'}^{\bm f} p_{n'_x}(t).
\label{master_eqn_nx}
\end{equation}

For the two-dimensional case considered in Sec.\ \ref{sec:2D}, the effective one-dimensional master equations,  from the point of view of an observer with access to only a single degree of freedom, either $x$ or $y$, respectively, are 
\begin{eqnarray}
\frac{dp_{n_x}(t)}{dt} &= & \left[ \kappa_{(1,0)}^{\bm 0}   e^{\alpha_{(1,0)} X_x} + \kappa_{(1,1)}^{\bm 0} e^{\alpha_{(1,1)}X_z}\right] p_{n_x-1}(t) -\Gamma_x p_{n_x}(t) \nonumber \\
& & \left[ \kappa_{(1,0)}^{\bm 0} e^{(\alpha_{(1,0)}-1)X_x} + \kappa_{(1,1)}^{\bm 0} e^{(\alpha_{(1,1)}-1)X_z}\right] p_{n_x+1}(t)  \label{mex}\\
\frac{dp_{n_y}(t)}{dt} &  = & \left[ \kappa_{(0,1)}^{\bm 0} e^{\alpha_{(0,1)}X_y} + \kappa_{(1,1)}^{\bm 0} e^{\alpha_{(1,1)}X_z}\right] p_{n_y-1}(t)-\Gamma_y p_{n_y}(t) \nonumber \\
& & \left[ \kappa_{(0,1)}^{\bm 0} e^{(\alpha_{(0,1)}-1)X_y} + \kappa_{(1,1)}^{\bm 0} e^{(\alpha_{(1,1)}-1)X_z}\right] p_{n_y+1}(t)  ,\label{mey}
\end{eqnarray}
where 
\begin{equation}
\Gamma_j = 2D_{jj}.
\label{D_jj}
\end{equation}
Coupled transitions $\kappa_{(1,1)}^{\bm 0}$ occur in both $x$ and $y$ simultaneously and are observed in both $x$ and $y$, whereas leak transitions $\kappa_{(1,0)}^{\bm 0}$ are observed in $x$ but not in $y$, and leak transitions $\kappa_{(0,1)}^{\bm 0}$ are observed in $y$ but not in $x$.  

The probability distribution for each coordinate can be determined analytically in the near and far from equilibrium limits.  Near equilibrium, i.e., when $|X_j|\ll 1$, the force dependence of the transition rates drops out and the solution to the master equations (\ref{mex}) and (\ref{mey}) with the initial state $p_{n_j}(0)=\delta_{n_j 0}$ is
\begin{equation}
p_{n_j}(t) = e^{-\Gamma_j t} I_{n_j} ( \Gamma_x t) ,
\end{equation} 
where $\Gamma_j \approx \sum_{\bm n\in{\rm for}, n_j \neq 0}2\kappa_{\bm n}^{\bm 0}$.  In the long-time limit, the probability distribution for each coordinate takes the Gaussian form
\begin{eqnarray}
p_{n_j}(t) &\sim&\frac{e^{-n_j^2/2\Gamma_x t}}{\sqrt{2\pi\Gamma_j t}},
\label{p_gauss}
\end{eqnarray}
characteristic of a diffusion process.

To consider the far from equilibrium limit, we take  $X_x\gg1$, $X_y \ll 1$, and $X_z=X_x+X_y \gg1$.  In this case, the following transition rates in the one-dimensional master equations (\ref{mex}) and (\ref{mey}) can be neglected: $\kappa_{(1,0)}^{\bm 0}\exp((\alpha_{(1,0)}-1)X_x)$, $\kappa_{(0,1)}^{\bm 0}\exp(\alpha_{(0,1)}X_y)$, and $\kappa_{(1,1)}^{\bm 0}\exp((\alpha_{(1,1)}-1)X_z)$. The master equation for the driving $x$ coordinate  can then be solved for the initial state $p_{n_x}(0)=\delta_{n_x 0}$ to yield the Poisson distribution
\begin{equation}
p_{n_x}(t) = e^{-\Gamma_x t} \frac{(t\Gamma_x)^{n_x}}{n_x !},
\label{poisson}
\end{equation} 
with $\Gamma_x \approx \kappa_{(1,0)}^{\bm 0}\exp(\alpha_{(1,0)}X_x)+\kappa_{(1,1)}^{\bm 0}\exp(\alpha_{(1,1)}X_z)$.  The master equation for the driven coordinate $y$ can be solved in the long-time limit, following the method presented in Sec.\ \ref{sec:diffusion}, to yield the Gaussian form
\begin{equation}
p_{n_y}(t) \sim \frac{e^{-(n_y-v_yt)^2/2\Gamma_y t}}{\sqrt{2\pi\Gamma_y t}},
\label{gauss_shift}
\end{equation}
with effective drift 
\begin{equation}
v_y  \approx   -\kappa_{(0,1)}^{\bm 0}e^{(\alpha_{(0,1)}-1)X_y}+\kappa_{(1,1)}^{\bm 0}e^{\alpha_{(1,1)}X_z},
\end{equation}
and effective diffusion half the decay rate
\begin{equation}
\Gamma_y  \approx  \kappa_{(0,1)}^{\bm 0}e^{(\alpha_{(0,1)}-1)X_y}+\kappa_{(1,1)}^{\bm 0}e^{\alpha_{(1,1)}X_z}.
\end{equation}

The interpretation of Eqs.\ (\ref{p_gauss}), (\ref{poisson}), and (\ref{gauss_shift}) will be discussed further in the following section.

\section{Single Trajectories \label{sec:single}}

\begin{figure}
	\centering
		\includegraphics{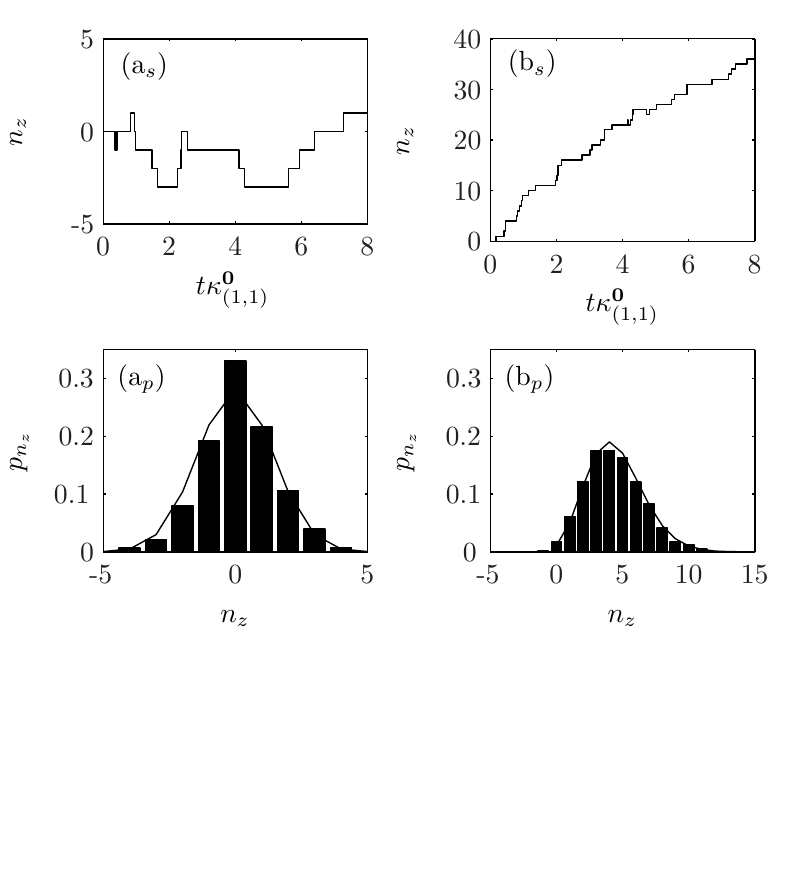}
		\vspace{-2.9cm}
		\caption{Single trajectories $(a_s)$ and $(b_s)$ and probability distributions $(a_p)$ and $(b_p)$ constructed from 1000 trajectories at $t\kappa_{(1,1)}^{\bm 0}=1$ with (a) $X_z = 0.1$ and (b) $X_z = 3$.  Other parameters are $\alpha_{(1,1)}=1/2$.  The curves are $(a_p)$ Eq.\ (\ref{soln_long_time}) and $(b_p)$ Eq.\ (\ref{poissonz}).}
	\label{fig:trajectory}
\end{figure}

Single trajectories of the system can be simulated using the decay rate $\Gamma$ and the hopping rates $\kappa_{\bm n}^{\bm f}$ \cite{Gardiner2009}.  We consider only the two-dimensional case.  In the strong-coupling regime, transitions occur along the coupled coordinate $z$ and single trajectories can be determined from the master equation (\ref{master_eqn_z}).  Figure \ref{fig:trajectory} shows single trajectories for (a) near and (b) far from equilibrium.  The probability distribution is determined numerically from an ensemble of single trajectories and compared with the analytic results from Sec.\ \ref{sec:strong}.  Near equilibrium, the rates of forward and backward hopping are comparable and the single trajectory is a random walk roughly balanced in the forward and backward directions.  The probability distribution is approximately Gaussian.  Far from equilibrium, the decay rate $\Gamma_z$ is larger than in the near-equilibrium regime, reducing the average waiting time.  Forward hops dominate and the single trajectory is a one-sided random walk. In the far equilibrium limit, there are no backward hops at all and the system evolves as a pure birth process with a Poisson probability distribution.  In the biochemical literature this is referred to as a Poisson enzyme \cite{Svoboda1994}.

\begin{figure}
\centering
\includegraphics{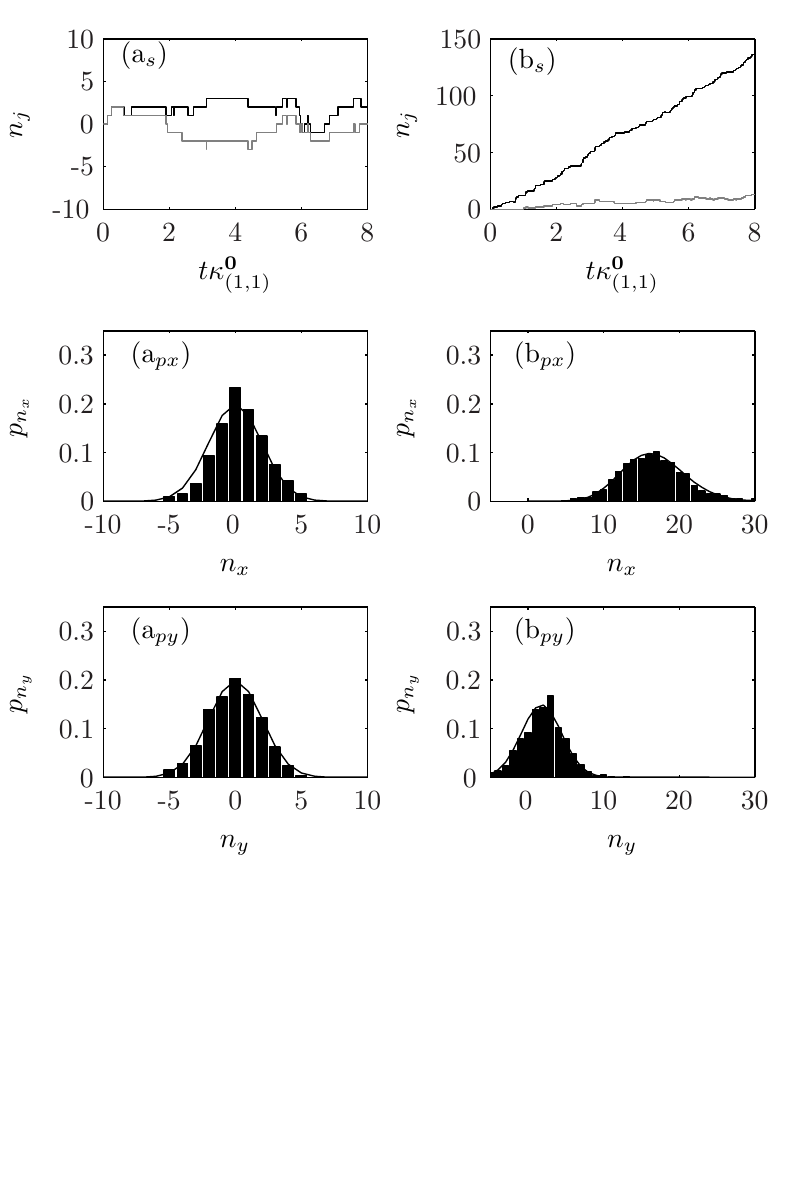}
\vspace{-3.6cm}
\caption{Single trajectories $(a_s)$ and $(b_s)$ in (black) $x$ and (gray) $y$, and probability distributions $(a_{px})$, $(a_{py})$, $(b_{px})$, and $(b_{py})$ constructed from 1000 trajectories at $t\kappa_{(1,1)}^{\bm 0}=1$.  Parameters are $(a)$ $X_x = 0.3$ and $X_y=-0.2$, and (b) $X_x = 5$ and $X_y = -2$. Other parameters are $\kappa_{(1,0)}^{\bm 0}=\kappa_{(0,1)}^{\bm 0}=\kappa_{1,1}^{\bm 0}$ and $\alpha_{(1,0)}=\alpha_{(0,1)}=\alpha_{(1,1)}=1/2$.  The curves are $(a_{px})$ and $(a_{py})$ Eq.\ (\ref{p_gauss}), $(b_{px})$ Eq.\ (\ref{poisson}), and $(b_{py})$ Eq.\ (\ref{gauss_shift}).}
\label{fig:traj_2d}
\end{figure}

In the weak-coupling regime, leak transitions are important and more than one degree of freedom must be described.  Figure \ref{fig:traj_2d} shows single trajectories for (a) near  and (b) far from equilibrium.  The probability distribution is determined numerically and compared with analytic results from Sec.\ \ref{sec:weak}.  Near equilibrium, the rates of forward and backward hops are comparable and a balanced random walk is observed in both the $x$ and $y$ coordinates.  The probability distribution is approximately Gaussian in both $x$ and $y$.  Far from equilibrium, the average waiting time is reduced and the ratio of forward to backward hopping rates biases the random walks.  In particular, the leak processes have a different effect on the driving and driven coordinates.  In the $x$ coordinate, forward hops dominate both for coupled and leak transitions and the probability distribution is approximately Poissonian.  In the $y$ coordinate, coupled transitions yield predominantly forward hops and leak transitions yield predominantly backward hops resulting in a Gaussian probability distribution shifted from the origin.  The driven coordinate only displays Poisson statistics in the limit of strong coupling.

Molecular motor experiments observe few, or no, backward steps in the driven mechanical coordinate \cite{Nishiyama2002, Yildiz2003, Shimabukuro2003, Toba2006, Ikezaki2013}.  The theoretical results presented here support the view that these motors operate, at least to a reasonable approximation, in the strongly-coupled, far-from-equilibrium regime.

\subsection{Drift and Diffusion}

Drift and diffusion can be determined from single trajectories by counting the number of forward and backward hops and determining the waiting time.  For example, if $N_j^f$ and $N_j^b$ are the number of forward and backward hops observed in coordinate $j$, then the drift in coordinate $j$ is
\begin{equation}
v_j = \Gamma_j \frac{N_j^f-N_j^b}{N_j^f+N_j^b}.
\end{equation}
For the diffusion matrix, the diagonal elements are given by the decay rates observed along the appropriate coordinate, as demonstrated by Eq.\ (\ref{D_jj}). The off-diagonal elements represent correlations between degrees of freedom and, to determine these from single trajectories, the relevant degrees of freedom must be observed simultaneously to identify hops occuring in coupled coordinates.  The simultaneous observation of mechanical and chemical coordinates has been demonstrated for myosin \cite{Yanagida2008}. In the two-dimensional case, if $N_z^f$ and $N_z^b$ are the number of forward and backward hops observed along the coupled $z$ coordinate, i.e., they are observed simultaneously in the $x$ and  $y$ directions, and $N_t$ is the total number of hops observed in all coordinates, then the off-diagonal diffusion matrix elements are given by
\begin{equation}
D_{xy}= D_{yx} = \Gamma \frac{N_z^f+N_z^{b}}{N_t}.
\label{Dxy_ob}
\end{equation}
An alternative approach is to determine the decay rate for hops only occuring in the coupled coordinate.  For either method, if the orthogonal coupling transition $\kappa_{(1,-1)}^{\bm 0}$ is non-negligible, care is needed to distinguish positive-x, positive-y transitions from orthogonal positive-x, negative-y transitions.  

\section{Conclusion \label{sec:conc}} 

We have used a Brownian theory for energy transfer in a molecular motor to derive formal expressions for the thermal fluctuation statistics.  Energy transfer between different degrees of freedom arises due to hopping transitions along coupled coordinates and leads to statistical correlations between thermal fluctuations in those degrees of freedom.  In the long-time limit, energy transfer can be described by a continuous diffusion process with a constant drift vector and difussion matrix.  The diffusion matrix quantifies the thermal fluctuations statistics in the steady state.

We have considered the discrete hopping statistics of the molecular motor and simulated single trajectories of the system.   Near equilibrium, single trajectories show a similar number of forward and backward hops and can be described by a random walk with negligible drift leading to Guassian statistics.  Far from equilibrium,  single trajectories are a one-sided random walk in the driving coordinate leading to Poisson statistics.  The driven coordinate undergoes a biased random walk, tending to a one-sided random walk only in the strong-coupling limit.

\appendix

\section{Gaussian Approximation Validity \label{sec:cumulant}}

The master equation (\ref{master_eqn}) can be derived by expanding the continuous Smoluchowski equation for overdamped Brownian motion on a multidimensional tilted periodic potential in the discrete Wannier basis of the untilted periodic potential \cite{Challis2013}.  In this description, the eigenvalue spectrum of the periodic potential displays a band structure with finite band gaps.  The master equation (\ref{master_eqn}) describes the lowest band and is valid for weak tilting and long times.  The weak-tilting condition ensures that coupling to higher bands is negligible.  The long-time condition ensures all higher bands have damped out, reducing the system dynamics to interwell hopping in the lowest band.  The long-time criterion for the master equation (\ref{master_eqn}) can be approximated by requiring all states above the lowest band to have damped out, i.e., we require $t \gg t_0$, where 
\begin{equation}
t_0 \sim \frac{1}{\max (Re\{\lambda_{\bm k}\})},
\label{t0_sba}
\end{equation}
and $\lambda_{\bm k}$ is the eigenvalue spectrum in the lowest band given by Eq.\ (\ref{eigenvalues_general}). 

The eigenstates $c_{\bm k}(t)$ of Eq.\ (\ref{characteristic_function_discrete}) can be interpreted as the characteristic function for the system and provide access to the system moments and cumulants \cite{Gardiner2009}. The eigenvalues $\lambda_{\bm k}$ can be expanded as a Taylor series around $\bm k = \bm 0$ yielding
\begin{equation}
\lambda_{\bm k} =- (-i)^d \left( \prod_{j'=1}^d \sum_{n_{j'}=0}^{\infty} \frac{a_{j'}^{n_{j'}} k_{j'}^{n_{j'}}}{n_{j'} !} \right) \langle\!\langle \dot{\prod_j n_j ^{n'_j} }\rangle \! \rangle _{ss},
\end{equation}
where $d$ is the number of coordinates and the expansion coefficients are the steady-state rate of change of the cumulants, i.e.,
\begin{eqnarray}
\langle\!\langle \dot{\prod_j n_j ^{n'_j} }\rangle \! \rangle _{ss} & = & \lim_{t\rightarrow \infty} \frac{\langle\!\langle {\prod_j n_j ^{n'_j} }\rangle \! \rangle}{t}\\
& =& -\prod_j \left. \left( \frac{i\partial}{a_{j}\partial k_{j}}\right)^{n'_{j}}\lambda_{\bm k} \right|_{\bm k = \bm 0} \\
& = & \sum_{\bm n \in {\rm for}} 2\kappa_{\bm n}^{\bm 0} G_{\bm n}\left(\frac{\bm X\cdot \bm n}{2}\right) F\left( \frac{\bm X \cdot \bm n}{2}\right) \prod_j n_j^{n'_j},
\end{eqnarray}
where $F(x) = \sinh (x)$ if $\sum_j n'_j$ is odd and $F(x) = \cosh (x)$ if $\sum_j n'_j$ is even.  The cumulants are convenient because they become less significiant with increasing order \cite{Gardiner2009}.  A consistent description is provided by the first two.  The first cumulant is related to the drift according to
\begin{eqnarray}
\langle\!\langle\dot{n_j}\rangle\!\rangle _{ss}& = &  v_j,
\end{eqnarray}
and the second cumulant is related to the diffusion matrix according to
\begin{eqnarray}
\langle\!\langle\dot{n_j n_{j'}}\rangle\!\rangle _{ss}& = &  2 D_{jj'}. 
\end{eqnarray}

The Gaussian approximation described in Sec.\ \ref{sec:diffusion} approximates the eigenvalue $\lambda_{\bm k}$ by its Taylor series around the origin truncated at second order.  This is equivalent to retaining only the first two cumulants and is valid for times long enough that  states lying outside the second-order truncation validity are damped out.  The long-time criterion can be estimated by considering the strong-coupling limit.  In this case, the second-order Taylor series approximation to the eigenvalue spectrum is accurate to within 10\% (20\%) for the real (imaginary) part of the eigenvalue spectrum for
\begin{equation}
  \sum_j k_j r_j = 1
  \label{sphere}
\end{equation}
On this sphere, the real part of the eigenvalue is down to $1/4$ of its maximum value within the lowest band.  Therefore, the long-time criterion for the Gaussian approximation is 
\begin{equation}
t \gtrsim 4 t_0,
\end{equation}
where $t_0$ is the timescale for the lowest-band approximation, given by Eq.\ (\ref{t0_sba}).



\end{document}